\documentclass{appolb}
\usepackage{epsfig}
\usepackage{url}
\usepackage{amssymb}

\begin{document}
\newcommand{\DAF}{DA$\Phi$NE}
\newcommand{\Reeps}{\ensuremath{\mathrm{Re}\,(\epsilon'/\epsilon)}}
\newcommand{\Imeps}{\ensuremath{\mathrm{Im}\,(\epsilon'/\epsilon)}}
\newcommand{\g}{\gamma}
\newcommand{\ee}{e^+e^-}
\title{The physics case of DAFNE-2
\thanks{Invited talk 
at the XXXI International Conference of Theoretical Physics 
``Matter To The Deepest'', Ustro\'n, 5-11 September 2007, Poland.}
\author{Graziano Venanzoni
\address{Laboratori Nazionali di Frascati, Frascati (RM) 00044, Italy}
\texttt{graziano.venanzoni@lnf.infn.it}
}}
\maketitle
\begin{abstract}

We present the physics case of DAFNE-2, an $e^+e^-$ collider
expected to deliver $20-50$ fb$^{-1}$ at the $\phi(1020)$ peak, 
and $\sim$ 5 fb$^{-1}$ in the energy region between 1 and 2.5 GeV.


\end{abstract}
\section{Introduction}
In the last decade a wide experimental program has been 
carried on at \DAF,  the 
$e^+e^-$ collider of the INFN Frascati National Laboratories 
running at a center of mass 
energy of 1020 MeV, the $\phi$ meson mass. 
Three experiments have run at \DAF:
KLOE, dedicated 
to kaon and hadronic physics, FINUDA, dedicated to the study of 
hypernuclei and DEAR, designed to study the production of kaonic 
atoms.
\par\noindent
In the last years a possible
continuation of a low energy $e^+e^-$ program has been considered. 
Two options 
emerged: (i) a continuation of the program at the 
$\phi$ peak with a luminosity significantly higher than the
present one (\DAF\ best peak luminosity was of 1.5$\times
10^{32}$cm$^{-2}$s$^{-1}$, corresponding to about 2 fb$^{-1}$ per year) 
and (ii) an increase of the 
\DAF\ energy up to at least 2.5 GeV. In the following we call DAFNE-2
the program based on both options.
While the second option seems 
technologically feasible, the first one is particularly challenging.
A new machine scheme (``Crabbed Waist'') 
aiming to increase the luminosity towards  10$^{33}$ cm$^{-2}$ s$^{-1}$
has been recently proposed by P. Raimondi, Head 
of the Frascati Accelerator Division~\cite{dafne2}. 
This scheme will be tested at \DAF\ in the next  months and it
will be used during the run of SIDDHARTA, an upgraded version of
DEAR experiment aiming to collect data in the 
first months of 2008. The
result of this machine test is very important in view also of
higher energy programs like 
the SuperB project. 

\section{DAFNE-2 Physics program}


Three Expressions of Interest (KLOE-2, AMADEUS, and DANTE) 
have been presented for DAFNE-2, with the following objectives:
\begin{itemize}
\item {\bf KLOE-2:} to continue the KLOE physics 
program, including tests of Quantum Mechanics and CPT
with kaon interferometry, measurement of the rare $K_S$ decays, 
test of lepton universality (as in the ratio $K_{e2}/K_{\mu2}$),
test of $\chi PT$ in the radiative decays of the $\phi$.
In addition, by using an electron tagger, precise measurements
of $\gamma \gamma$ physics can be performed.
The high energy program (from 1 to 2.5 GeV)
will allow a precise measurement of  the hadronic cross sections, and 
vector meson spectroscopy;
\item {\bf AMADEUS:} to study the deeply bound kaonic nuclear states by using 
gaseous targets around the interaction region;
\item {\bf DANTE:} to measure time-like form factors of nucleons and lower mass hyperons.
\end{itemize}
In the following we will discuss  
three arguments from the KLOE-2 EoI:
$(i)$ CPT tests with kaon interferometry;  
$(ii)$ the measurement of the hadronic cross section (R-measurement)
in the 1--2.5 GeV energy region\footnote{This topic  
is also considered in  DANTE EoI.};
$(iii)$ $\gamma\gamma$ physics.
For the whole DAFNE-2 program we refer the reader to~\cite{eoi,noi}.

\subsection{Kaon interferometry and CPT tests}
\begin{figure}[h]
\special{hoffset=-70 voffset=-300}
\centering
\epsfig{file=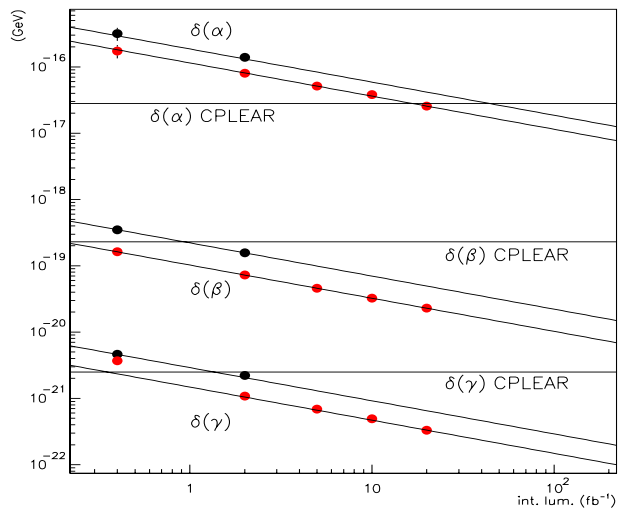,width=10cm,height=11cm}
\vspace{-2.cm}
\caption{Limits on the CPT violating parameters 
$\alpha$, $\beta$, and $\gamma$ obtainable by KLOE-2 as a function of the integrated
luminosity. Results are presented for a detector  
both with and without the insertion of an inner tracker
with vertex resolution of 0.25~$\tau_{S}$ (to be compared with the present
 KLOE vertex 
resolution, 0.9~$\tau_{S}$). In the figure also are given 
results from CPLEAR.}
\label{fig:qmp}
\end{figure}
$CPT$ invariance is a fundamental theorem in quantum field theory.
In several quantum gravity (QG) models, however, $CPT$ 
can be violated
via some  mechanism which can also 
violate standard Quantum Mechanics (QM). 
In this respect the entagled neutral kaon 
pairs produced at \DAF\  play an unique role
in precision tests of the $CPT$ symmetry~\cite{bemp}.
As an example of this incredible precision reachable with neutral kaons,
let's consider the model by Ellis, Hagelin, Nanopoulos and Srednicki (EHNS)
 which introduces three $CPT$ and QM-violating
real parameters $\alpha$ $\beta$ and $\gamma$~\cite{ehns}.  
On phenomenological grounds, they are expected to 
be \break
 $O(m^{2}_{K}/M_{Pl})\sim2\times$10$^{-20}$GeV at most, since 
$M_{Pl}\sim10^{19}$ GeV, the Planck mass. 
Interestingly enough, this model gives
 rise to observable effects in the behaviour
 of entagled neutral meson systems, as shown also in \cite{phu},  
 that can be experimentally tested.
KLOE has already published competitive results on these issues, 
based on a statistics of $\sim400$ pb$^{-1}$~\cite{kqm}.
The analysis makes use of correlated $K^{0}_{L}-K^{0}_{S}$ pairs,  
by measuring the relative distance of their decay point into two charged pions. 
The decay region most sensitive  to the EHNS parameters is the one close to
the IP. 
 
Fig.~\ref{fig:qmp} shows the potential limits that can be obtained by KLOE on 
$\alpha$, $\beta$, and $\gamma$ as a function of the integrated luminosity, 
both with and without the insertion of an inner tracker (see Sect.~\ref{dec})
with vertex resolution of 0.25~$\tau_{S}$ (to be compared with the present
 KLOE vertex 
resolution, 0.9~$\tau_{S}$). In the figure also are given the  
results from CPLEAR~\cite{cplearqm}.
Without entering too much in details, it is clear that with a 
reasonable  integrated luminosity, KLOE-2 can set the best limits on these 
parameters.

\subsection{Measurement of R  in the 1--2.5 GeV energy region.}

The ratio $R=\frac{\sigma(e^+e^-\to hadrons)}{\sigma(e^+e^-\to\mu^+\mu^-)}$ 
is poorly known in the region [1--2.5 GeV], where the uncertainty 
is $\sim15\%$. This region contributes to about  40\% 
to the total error on the dispersion integral for 
$\Delta^{(5)}_{had}(m_Z^2)$~\cite{noi}.
 It also provides
most of the contribution to $a_{\mu}^{\mbox{$\scriptscriptstyle{\rm HLO}$}}$ above 1 
GeV~\cite{noi,fj}. Recently~\cite{fj0} a new approach has been proposed to evaluate 
$\Delta^{(5)}_{had}(m_Z^2)$.
Based on the evaluation of the so called Adler function, 
it allows to use safetely pQCD down to 2.5 GeV {\it plus}
 experimental data up to that threshold. 
In this approach DAFNE-2 can play a substantial role, and a measurement of R at 1\% level 
below 2.5 GeV can considerably improve the accuracy
 on $\Delta^{(5)}_{had}(m_Z^2)$~\cite{fj0}.
As an example of the statistical accuracy that can be reached in this region
we will consider the process $e^+ e^- \to$ 3 and 4 hadrons.
\begin{figure}[h]
\begin{center}
\epsfig{figure=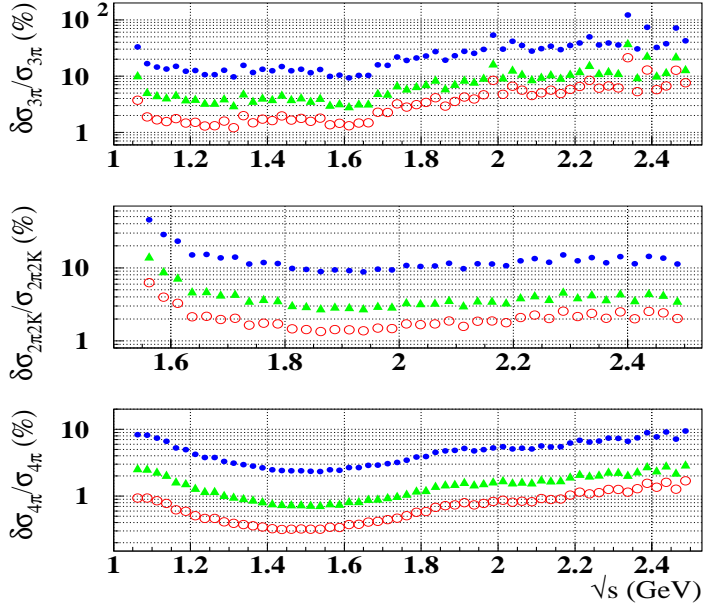,width=10cm,height=10cm}
\vspace{-1.cm}
\caption{\label{fig:impactscan} Comparison of the statistical accuracy in
the cross-section among DAFNE-2 with an energy scan with 20 pb$^{-1}$ per 
point ($\circ$); published {\small BABAR} results ($\bullet$);
{\small BABAR} with full statistics ($\blacktriangle$);  for $\pi^+\pi^-\pi^0$
(top), $\pi^+\pi^-K^+K^-$ (middle) and $2\pi^+ 2\pi^-$ (down) channels. 
An energy step of 25 MeV is assumed.}
\end{center}
\end{figure}

 {\small BABAR} has already published results on these channels,
 obtained with an integrated luminosity 
of 89 fb$^{-1}$,
and it is expected to reach 1 ab$^{-1}$ by the end of the data taking.
However, due to the ISR photon emission at the $\Upsilon(4s)$ resonance, 
the effective luminosity for tagged photon ($\theta_{\gamma}>20^o$)
for energies below 2.5 GeV, will be of the order of few pb$^{-1}$ 
at full statistics~\cite{noi}. 
%
%
%
%
%
Fig.~\ref{fig:impactscan}  shows the statistical error for the channels
$\pi^+\pi^-\pi^0$, $2\pi^+ 2\pi^-$ and $\pi^+\pi^-K^+K^-$, which can be
 achieved by  an energy scan at DANFE-2 with 20 pb$^{-1}$ per point,
compared  
with {\small BABAR} with published (89 fb$^{-1}$), and 
expected full  (890 fb$^{-1}$) statistics.
As it can be seen, an energy scan allows to reach a statistical
 accuracy of the order of  1\% on these channels.

\subsection{$\gamma\gamma$ physics}
The term ``$\gamma\gamma$ physics'' (or ``two-photon physics'') stands for the 
study of the reaction:
$$
\ee\,\to\,\ee \,\g^*\g^*\,\,\to\,\ee \,+\, X
$$
where $X$ is some arbitrary final state allowed by conservations laws. 

\begin{figure}[ht]
\centering
\mbox{\epsfig{file=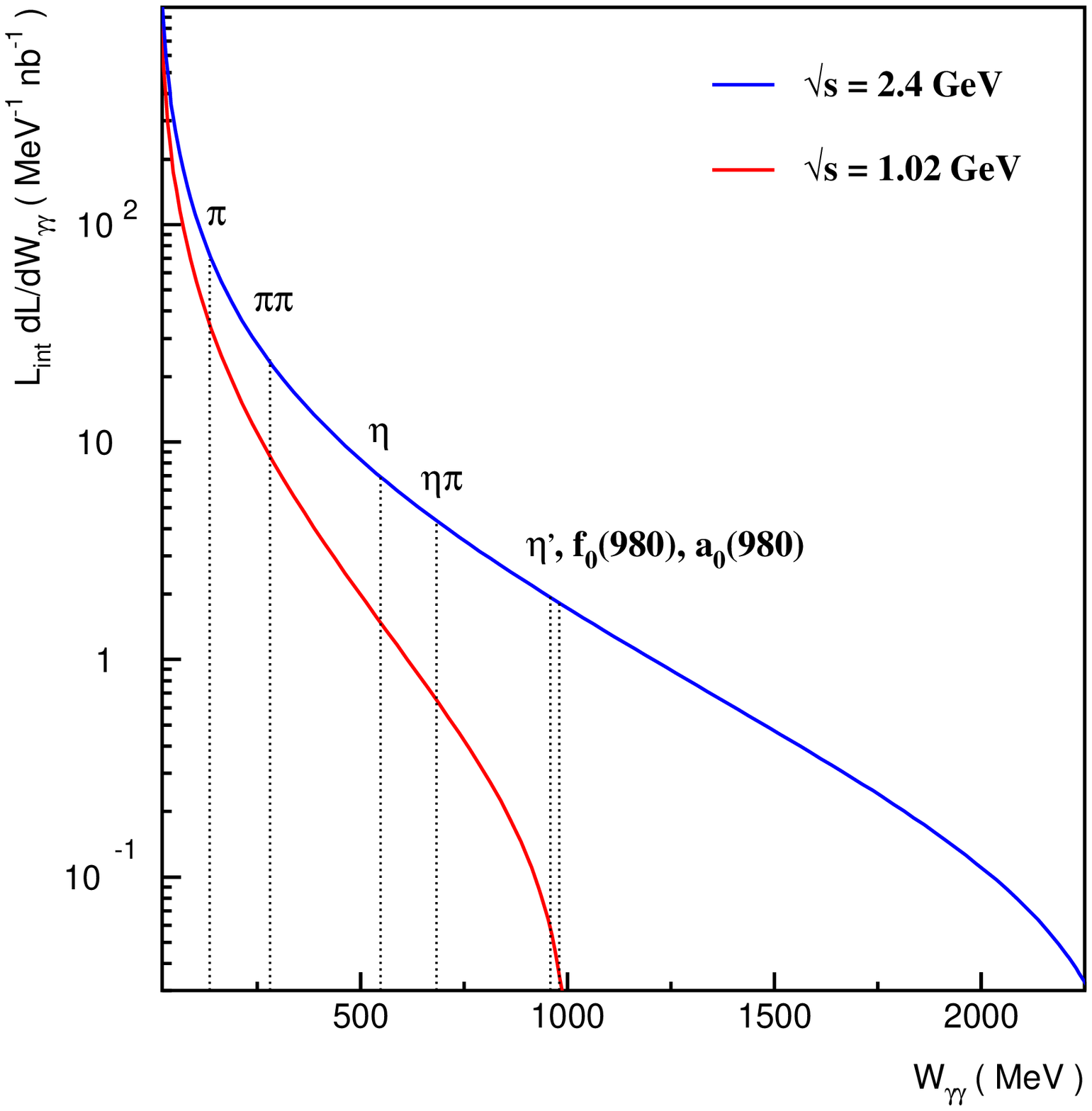,width=6cm,height=6cm}
\epsfig{file=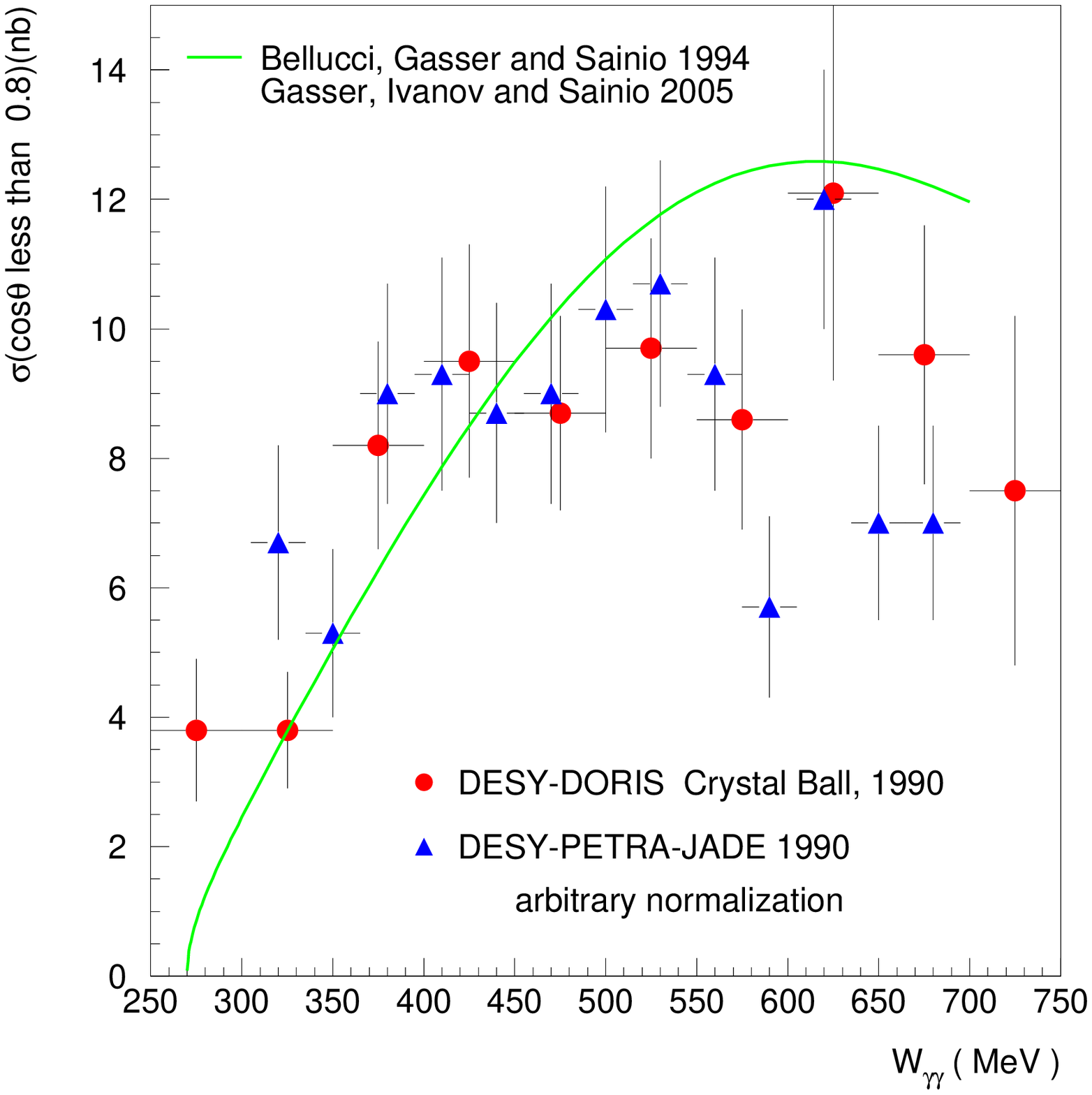,width=6cm,height=6cm}}
\caption{\small{{\it Left}: Effective $\gamma\gamma$ luminosity as a function of
 $W_{\gamma\gamma}$ corresponding to an integrated luminosity 
 of  1 fb$^{-1}$ at $\sqrt{s} = m_{\phi}$ (red curve) and at 
$\sqrt{s}$=2.4 GeV (blue curve). Vertical lines represent from left 
to right: $\pi$-threshold, $\pi \pi$-threshold, $\eta$, $\eta \pi$-threshold,
$\eta'$, $f_0$, $a_0$. {\it Right}: Collection of low energy $\gamma\gamma\rightarrow\pi^{0} \pi^{0}$ 
                cross-section data compared with a theoretical evaluation  		    based on $\chi$PT \cite{Bellucci}. The JADE data are normalised 		    to the same average cross-section of the Crystal Ball data. 
}}
\label{gg}
\end{figure}

The number of $e^+e^-\rightarrow e^+e^-X$ events per unit of invariant mass $W_{\gamma\gamma}$, as a function 
of $W_{\gamma\gamma}$ itself, is:
\begin{eqnarray*}
N({\rm evts/MeV})& = & L_{\rm int}({\rm nb^{-1}}) \times \\
\frac{{\rm d} F (W_{\gamma \gamma},\sqrt{s})}{{\rm d} W_{\gamma\gamma}}
({\rm MeV^{-1}}) 
& \times & \sigma ( \gamma \gamma \rightarrow X) ({\rm nb})
\end{eqnarray*}
where $L_{\rm int}$ is the $e^+e^-$ integrated luminosity and 
${\rm d} F(W_{\gamma\gamma},\sqrt{s})/{\rm d} W_{\gamma\gamma}$ is the effective 
$\gamma\gamma$ luminosity per unit energy. The product ${\rm d} F/{\rm d} W \times L_{\rm int}$ 
is reported in Fig.~\ref{gg} ({\it Left}) for two DAFNE-2 center of mass (c.m.)
 energies.

\subsubsection{The process $\gamma\gamma \rightarrow \pi^o\pi^o$: the $\sigma$ case }
$\gamma\gamma$-physics provides a complementary
 view at the light scalar mesons and, in 
particular, is a powerful tool to search for the
 $\sigma$. $e^+e^-\rightarrow e^+e^-X$ events 
with $X = \pi\pi$, $\eta\pi$ and possibly $K\bar{K}$, 
allow to study directly the $I = 0$ 
and $I = 1$ scalar amplitudes down to their thresholds. 
In $\gamma\gamma \rightarrow \pi^0\pi^0$ 
events with two-photon invariant masses $W_{\gamma\gamma}$ below 1 GeV, 
the $\pi^0\pi^0$ pair is mostly in s-wave, resulting in $J^{PC} = 0^{++}$
 quantum numbers, with negligible contamination 
from other hadronic processes.
 The presence of a pole in this amplitude around 500 MeV 
\cite{Caprini} would be a clean and new signal of the $\sigma$.  
 
Unfortunately, the only available  experimental information on this 
channel in the region of interest is relatively poor
and do not allow to draw any conclusion about the agreement with either the 
$\chi$PT
calculations nor on the  existence of the broad (250-500 MeV) 
$\sigma$ resonance  (see Fig.~\ref{gg} ({\it Right})). 
A new measurement of  $\gamma\gamma \rightarrow \pi^0\pi^0$ in this 
region would be therefore very important~\cite{nguyen}.


\subsubsection{Measurement of the $\gamma\gamma$ widths of $f_0(980)$ and $a_0(980)$} 
Extending the measurement of $\gamma\gamma \rightarrow \pi\pi$ and 
$\gamma\gamma \rightarrow \eta\pi$ up to $W_{\gamma\gamma}\,\sim\,1$ GeV, the two-photon 
width of $f_0(980)$ and $a_0(980)$ can also be measured. This measurement is possible by 
running at the maximum attainable centre of mass energy, in order to maximise
the effective $\gamma\gamma$ luminosity in the GeV region 
(see Fig.~\ref{gg}, {\it Left}).
In both 
cases a peak in the $W_{\gamma\gamma}$ dependence of the 
$\gamma\gamma \rightarrow \pi\pi (\eta\pi)$ cross-section around the meson mass allows to 
extract the $\gamma\gamma$-width.

\subsubsection{The two-photon widths of the pseudoscalar mesons}
The
situation on the decay constants of  $\eta$ and $\eta'$
is far from being satisfactory and calls for more precise measurements 
of the two-photon width of these mesons~\cite{noi}. Even the
 $\pi^0$ two-photon width is poorly known (relative
uncertainty of $\sim 8\%$) and its determination can be
improved at DAFNE-2. Given the small value 
of these widths, the only way to measure them is the
 meson formation in $\g\g$ reactions. 
In Tab.\ref{ggtab} we report the estimates for the total production rate of 
a pseudoscalar meson (PS) in the process $e^+e^-\rightarrow e^+e^-PS$ 
for two DAFNE-2 c.m. energies~\cite{noi}. 
\begin{table}[ht]
  \centering
  \begin{tabular}{| c | c | c | c |}
    \hline
    $\sqrt{s}$ (GeV) & $\pi^0$ & $\eta$ & $\eta'$ \\
    \hline
    1.02 & 4.1$\times 10^5$ & 1.2$\times 10^5$ & 1.9$\times 10^4$ \\
    2.4 & 7.3$\times 10^5$ & 3.7$\times 10^5$ & 3.6$\times 10^5$ \\
    \hline
  \end{tabular}
  \caption{$e^+e^-\rightarrow e^+e^-PS$ total rate for an integrated
    luminosity of 1 fb$^{-1}$ at two different center of mass energies. No
    tag efficiency is included in the rate calculation.}
  \protect\label{ggtab}
\end{table}

\subsubsection{Meson transition form factors}\label{sec:ffact} 
The process $e^+e^- \to e^+e^- + PS$ with one of the final 
leptons scattered at large angle gives access to the process 
$\g\g^* \to PS$, i.e. with one off-shell photon, and it allows to extract
information on the pseudoscalar meson transition form factor $F_{P\g\g^*}(Q^2)$.
A precise determination of this quantity would be important 
to test phenomenological models. 

By detecting both the leptons at large angle, the doubly off-shell 
form factor $F_{P\g^*\g^*}(Q^2_1,Q^2_2)$ can be accessed.
A direct and accurate determination of this quantity would be 
extremely important in order to get less model-dependent estimations of 
the hadronic  light-by-light scattering 
in $(g-2)_{\mu}$~\cite{fj}.
%

\section{Detector consideration}
\label{dec} 
The KLOE detector is well suited for most 
of the measurement that can be carried out with DAFNE-2. However
some upgrades are expected~\cite{eoi}: 
\begin{itemize}
\item[-] An inner tracker in the region between the beam
 pipe and the inner wall 
of the drift chamber, which is presently not instrumented;
\item[-] The equipment of the electromagnetic calorimeter with photomultipliers
 with higher 
quantum efficiency;
\item[-] New quadrupole calorimeters (QCAL);
\item[-] Electron taggers, 
needed for $\gamma\gamma$ physics. 

\end{itemize}

The measurement of the nucleon form factors with KLOE can be more problematic, 
since a proton polarimeter is required. Such
a device normally consists of a layer of carbon placed between two precise
tracking devices, typically silicon detectors. This object cannot be easily
incorporated in the {\small KLOE} structure and would spoil the tracking resolution
of the detector. It should then be inserted only for a dedicated
run,  maybe replacing part of the beam pipe or of the vertex detector.
Finally the wide program of measurements of the $KN$ interactions in the
$p_K\sim100$ MeV/c momentum region, requires different 
gaseous targets around the interaction region~\cite{eoi}.

\section{Conclusion}
The physics program of DAFNE-2
has been discussed. 
It is a wide physics program, 
based on the possibility to increase the
luminosity at the $\phi$(1020) peak and to extend the center
of mass energy up to 2.5 GeV. Such a machine will allow to perform
fundamental tests of CPT symmetry and QM, precision tests of the Standard Model, and 
a large number of relevant 
measurements in the hadronic sector.

\vspace{.5 cm}

\section*{Acknowledgements} 

I thank my colleagues from KLOE-2, AMADEUS, and DANTE for 
useful discussions. I'm particularly grateful to 
D. Babusci, C. Bini, C. Bloise, F. Bossi, 
S. Eidelman, F. Jegerlehner and M. Zobov for numerous clarifications.



\begin{thebibliography}{9}
\bibitem{dafne2} P. Raimondi, 2$^{nd}$ SUPERB Workshop, Frascati 2006;
  P.~Raimondi, D.~N.~Shatilov and M.~Zobov,
  arXiv:physics/0702033; D. Alesini {\it et al.} LNF-06/33 (IR) (2006).
\bibitem{eoi}   \url{http://www.lnf.infn.it/lnfadmin/direzione/roadmap/KLOE2-LoI.pdf};
\url{http://www.lnf.infn.it/lnfadmin/direzione/roadmap/LOI_MARCH_AMADEUS.pdf};
 \url{http://www.lnf.infn.it/conference/nucleon05/FF}.
\bibitem{noi} F. Ambrosino {\it et al.}, 
Eur.\ Phys.\ J.\  C {\bf 50} (2007) 729.
\bibitem{bemp}J.~Bernabeu, J.~Ellis, N.E.~Mavromatos, D.V.~Nanopoulos and 
J.~Papavassiliou, arXiv:hep-ph/0607322.
\bibitem{ehns} J.~Ellis, J.S.~Hagelin, D.V.~Nanopoulos, M.~Srednicki, 
Nucl. Phys. B {\bf 241}, 381 (1984).
\bibitem{phu} P.~Huet, M.~Peskin, Nucl. Phys. B {\bf 434}, 3 (1995).
\bibitem{kqm} F.~Ambrosino et al. (KLOE Collaboration)
Phys. Lett. B {\bf 642}, 315 (2006).
\bibitem{cplearqm} A.~Adler et al. (CPLEAR Collaboration), 
Phys. Lett. B {\bf 364} 239 (1995).
\bibitem{fj} F. Jegerlehner, arXiv:hep-ph/0703125.
\bibitem{fj0} F. Jegerlehner,  Nucl. Phys. B Proc. Suppl. {\bf 162}, 22 (2006).
\bibitem{Caprini}
I.~Caprini, G.~Colangelo and H.~Leutwyler,
  Phys.\ Rev.\ Lett.\  {\bf 96} (2006) 132001.
\bibitem{nguyen} F.~Nguyen, F.~Piccinini and A.~D.~Polosa,
  Eur.\ Phys.\ J.\  C {\bf 47} (2006) 65.
\bibitem{Bellucci}  
  S. Bellucci {\em et al.},  
  Nucl. Phys. B{\bf 423}, 80 (1994); erratum-ibid. B {\bf 431}, 413 (1994); 
  J. Gasser,  {\em et al.},  
  Nucl. Phys. B {\bf 728}, 31 (2005).  
\end{thebibliography}
\end{document}